\documentclass[aps,prb,twocolumn,superscriptaddress,showkeys,showpacs]{revtex4}
\usepackage{amssymb,latexsym,amsthm}
\usepackage[dvips]{graphics}

\def\a{\alpha}
\def\b{\beta}
\def\g{\gamma}
\def\d{\delta}
\def\e{\epsilon}
\def\s{\sigma}

\def\D{\Delta}

\def\L{\Lambda}

\def\r{\rho}
\def\C{{\cal C}}

\theoremstyle{plain}
\newtheorem{thm}{THEOREM}[section]

\newtheorem{prop}[thm]{PROPOSITION}
\theoremstyle{definition}
\newtheorem{defin}[thm]{DEFINITION}
\theoremstyle{remark}
\newtheorem{remark}[thm]{REMARK}
\newtheorem{remarks}[thm]{REMARKS}
\newtheorem{assumption}[thm]{ASSUMPTION}

\newcommand{\be}{\begin{equation}}
\newcommand{\ee}{\end{equation}}
\newcommand{\bea}{\begin{eqnarray}}
\newcommand{\eea}{\end{eqnarray}}
\newcommand{\beax}{\begin{eqnarray*}}
\newcommand{\eeax}{\end{eqnarray*}}

\newcommand{\mfr}[2]{{\textstyle\frac{#1}{#2}}}

\begin{document}

\preprint{UC Davis Math 2002-07}

\title{The Ferromagnetic Heisenberg {\sf XXZ} chain in a pinning field}


\author{Pierluigi Contucci}
\email[]{contucci@dm.unibo.it}
\affiliation{Dipartimento de Matematica \\
Universit\`a di Bologna \\
40127 Bologna, Italy}
\author{Bruno Nachtergaele}
\email[]{bxn@math.ucdavis.edu}
\author{Wolfgang L.~Spitzer}
\email[]{spitzer@math.ucdavis.edu}
\affiliation{Department of Mathematics \\
\normalsize University of California \\
\normalsize Davis, CA 95616-8633 USA}

\date{\today}

\begin{abstract}
We investigate the effect of a magnetic field supported at a
single lattice site on the low-energy spectrum of the
ferromagnetic Heisenberg {\sf XXZ} chain.  Such fields, caused by
impurities, can modify the low-energy spectrum significantly by
pinning certain excitations, such as kink and droplet states. We
distinguish between different boundary conditions  (or sectors),
the direction and also the strength of the magnetic field. E.g.,
with a magnetic field in the $z$-direction applied at the origin
and $++$ boundary conditions, there is a critical field strength
$B_c$ (which depends on the anisotropy of the Hamiltonian  and the
spin value) with the following properties: for $B < B_c$ there is a
unique ground state with a gap, at the  critical value, $B_c$,
there are infinitely many (droplet) ground states with gapless
excitations, and for $B>B_c$ there is again a unique ground state
but now belonging to the continuous spectrum. In contrast, any
magnetic field with a non-vanishing component in the $xy$-plane
yields a unique ground state, which, depending on the boundary
conditions, is either an (anti)kink, or an (anti)droplet state.
For such fields, i.e., {\it not} aligned with the $z$-axis,  excitations
always have a gap and we obtain a rigorous lower bound for that
gap.
\end{abstract}

\pacs{75.10.-b, 
75.60.-d, 
05.50.+q 
}
\keywords{Quantum spin model, Heisenberg {\sf XXZ} chain, kink states, 
domain wall, spectral gap}

\maketitle

\section{Introduction}

The quantum spin-$j$ Heisenberg {\sf XXZ} chain has been the focus
of intensive studies in recent years. The spin 1/2 chain by itself
has connections with a surprising variety of interesting
mathematical structures, such as quantum groups \cite{PS}, vertex
algebras \cite{JM}, and fundamental  problems in combinatorics
\cite{Kup,BdGN}, to name just a few. Of course, the interest in
the {\sf XXZ} model is not limited to mathematics. In 1995,
Alcaraz, Salinas, and Wreszinski \cite{ASW} and, independently,
Gottstein and Werner \cite {GW} discovered that, with suitable
boundary terms, the ferromagnetic {\sf XXZ} chain possesses a
family of kink ground states which describe a domain wall of
finite thickness (these domain walls are exponentially localized,
with a width depending on the anisotropy parameter $\Delta$, which
diverges as $\Delta\downarrow 1$). Moreover, it was shown in
\cite{ASW}, that similar states exist for the spin $j$ model for
arbitrary $j$ and in all dimensions.

From the physical point of view, the discovery of Giant
Magnetoresistance and its connection with transport properties in
the presense of magnetic domain walls has also  spurred renewed
interest in the microscopic description of domain walls
\cite{DCS,GBB,JPRT,KY1,KY2,KY3,MNSO,NN,TF,WCJADN}. Of particular
relevance are low-lying excitations associated with  them. Koma
and Nachtergaele discovered that, although the {\sf XXZ} model has
a gap in its spectrum above the trivial translation invariant
ground  states, gapless excitations exists associated with
diagonal domain walls (11, 111, ...) in two or more dimensions
\cite{KN2}. The scaling behavior of these excitations was recently
determined in \cite{BCNS} and \cite{CapMart1}, \cite{CapMart2}.

It is interesting to note that the kink and antikink ground states
were discovered by a careful study of the {\sf XXZ} chain with the
special boundary conditions that make the spin 1/2 model
$SU_q(2)$-symmetric. Although this quantum group symmetry is
destroyed for $j>1/2$ or $d>1$, interface ground states exist in
general. In one dimension it has been proven rigorously that no
other ground states exist, in the sense of local stability, no
matter what boundary conditions are considered \cite{Mat,KN3}. In
the last reference it was also proved that the {\sf XXX} chain
does not have domain wall ground states that are stable in the
infinite-volume limit.

There is an obvious need for a clear understanding of excitations near
magnetic interfaces in order to develop more accurate models of electron
scattering at such interfaces. It has been noted, however,  that pinning of
interfaces by impurities may have to be taken into account  as well
\cite{JPRT}. Here, we study the {\sf XXZ} chain perturbed at one site by a
magnetic field as a caricature model for a pinned domain wall. Admittedly,
the one-dimensional nature of the model restricts its direct applicability to
experimental situations. We will see however that the low-lying spectrum
of this model of pinned interfaces exhibits a number of interesting features
that we expect will carry over, {\em mutatis mutandis}, to the two- and
three-dimensional case.

Let us now define the model precisely and briefly summarize our main results. 
The spin $j$ {\sf XXZ} Hamiltonian without
boundary terms and with anisotropy parameter $\D>1$ is defined on
the finite chain labeled by the integers from $a$ to $b$ by
 \be\label{H_0 bare}
    H_0^{} = -\sum_{x=a}^{b-1} \left[\frac{1}{\D}(S^1_x S^1_{x+1}
    + S^2_x S^2_{x+1})+ S^3_x S^3_{x+1} - j^2{\bf 1}\right]  ,
 \ee where the copies of the spin operators at position $x$,
$S_x^1,S_x^2,S_x^3$ satisfy the usual commutation relations:
 \be
 [S_x^\a,S_x^\b]= i\,\e^{\a \b \g}\,S_x^\g .
 \ee
The anisotropy $\D>1$ has been put in front of the ${\sf XX}$ part
so that we can easily take the Ising limit $\D\to\infty$. Unless
otherwise stated, we set $a=1$.
We consider the perturbation of $H_0$ obtained by adding a term
$\vec{B}\cdot \vec{S}_y$, i.e. a magnetic field at the site $y$.
As a way to impose boundary conditions, we also add
magnetic fields in the $z$-direction at the boundary spins. First,
consider boundary fields in the negative $z$-direction at both
ends, which we will refer to as ($++$) boundary conditions (b.c.),
indicating that they favor the spins at the ends to point in the
positive $z$-direction.

If the perturbation at the interior site $y$ has a component
orthogonal to the $z$-direction, we find that the ground state is
unique and describes a {\em droplet state}, i.e., the
magnetization is reduced from its maximum possible value in some
neighborhood of $y$. Strictly speaking, the magnetization is
reduced everywhere in the chain, but by an amount that decays
exponentially fast away from $y$.

\begin{figure}[h]
\resizebox{3in}{2in}{\includegraphics{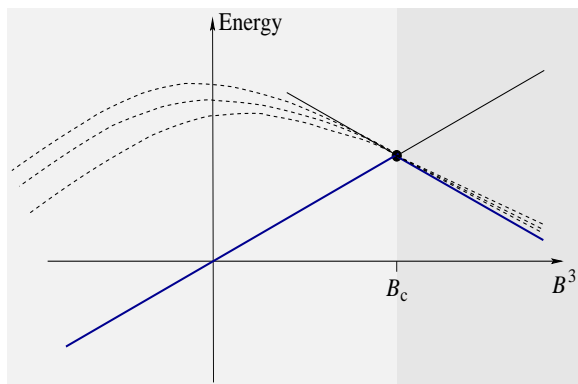}}
\caption{\label{fig1} Ground state energy (thick line) in a
magnetic field in the $z$-direction and $(++)$ or $(--)$ b.c.. The
dotted lines indicate droplet states which describe excitations
except at one value of the field strength, $B_c$, given in Proposition
\ref{droplet sector}. Also see Figure \ref{fig5}.}
\end{figure}

However, when the field is in the $z$-direction, then there is a
critical value $B_c$ such that for $B<B_c$, the all spin-up state
is the ground state. At $B=B_c$, there are infinitely many ground
states which are droplet states describing domains of negative
magnetization of arbitrary size embedded in a environment of
positive magnetization. For stronger fields, $B>B_c$, the magnetic
field selects the all spin-down state as its ground state. This is
illustrated in Figure \ref{fig1}.

The ground state picture is simpler when we impose ($+-$) b.c.,
i.e. fields in opposite directions at the boundary spins. For
boundary fields of magnitude (\ref{DefA}), and without a
perturbation in the interior, we then have a set of {\em kink
states} as the ground state, one for every possible value of the
magnetization \cite{ASW,GW}. In that case, any non-zero field
$\vec{B}$ at an interior site $y$ selects a unique ground state.
If $\vec{B}$ is parallel to the $z$-direction the ground state is
in the continuous spectrum, and hence there are excitations of
arbitrary small energy. If there is a non-vanishing component of
$\vec{B}$ in the $xy$-plane, the unique ground state is separated
by a gap form the rest of the spectrum. The unique ground state is
a kink state centered at a position which we calculate. We also
obtain an estimate for the gap.

In Section 2, we define the model and find the set of ground
states. Section 3 is devoted to the study of the gap in the spin
$1/2$ case. Some less illuminating calculations are presented in
three appendices.

\section{The model and its ground states}

The main lesson to be learned from the proof of completeness of
the list of ground states of the infinite ferromagnetic XXZ chain
(cf.~\cite{KN1}), is that one only needs to study finite chain
Hamiltonians with very simple b.c. This remains true if we add a
bounded perturbation with finite support to the Hamiltonian, e.g.,
a magnetic field at one site. These simple b.c. are fields in the
$z$-direction with either equal or opposite sign which we will
introduce shortly. 

The Hamiltonian $H_0$, defined in (\ref{H_0 bare}), is non-negative, and the two
translation invariant all spin-up/down states are the ground states of $H_0$.
It will be convenient to separate them by adding the equal-sign boundary
fields, $jA(S_1^3+S_b^3)$ with
\be\label{DefA} A = A(\D)= \sqrt{1-\D^{-2}}, 
\ee and define the {\it droplet} and {\it antidroplet} Hamiltonian
\bea
   H_0^{++}& = &H_0^{} - jA(S^3_{1} + S^3_{b} - 2j{\bf 1}) ,
 \\\label{} \nonumber
 H_0^{--} &=& H_0^{} + jA(S^3_{1} + S^3_{b} + 2j{\bf 1}) .
 \eea
For convenience we have normalized the ground state energy to 0.
By reflecting all $S^3_x$ into$-S^3_x$ the two Hamiltonians,
$H^{++}$ and $H^{--}$ are unitarily equivalent and we only
study $H_0^{++}$.

Additional ground states emerge when we add opposite-sign boundary
terms. It turns out that precisely for the fields $\pm jA(S^3_{1}
- S^3_{b})$ one discovers the full set of new ground states.
Therefore, we define the {\it kink} and {\it antikink} Hamiltonian
 \bea\label{H_0 kink}
 H_0^{+-}& = &H_0^{} - jA(S^3_{1} - S^3_{b}) ,
 \\ \label{H_0 antikink}
 H_0^{-+} &=& H_0^{} + jA(S^3_{1} - S^3_{b}) .
 \eea
Again, by spin reflection, the kink and anti-kink Hamiltonians are
unitarily equivalent. Let us define
 \bea \label{h^{+-}}
 h_{x x+1}^{+-} &=&-\mfr{1}{\D}(S^1_x S^1_{x+1} + S^2_x S^2_{x+1}) -
 S^3_x S^3_{x+1} + j^2 {\bf 1} \nonumber
 \\
 &&- j A(S^3_{x} - S^3_{x+1}) ,
 \eea
and
 \bea \label{h^{++}} h_{x x+1}^{-+}
 &=&-\mfr{1}{\D}(S^1_x S^1_{x+1} + S^2_x S^2_{x+1})
 - S^3_x S^3_{x+1} + j^2 {\bf 1}
 \nonumber
 \\
 &&+j A(S^3_{x} - S^3_{x+1}) .
 \eea
In terms of these interactions terms we may write
 \be
 H_0^{+-} = \sum_{x=1}^{b-1} h_{x x+1}^{+-},
 \ee
and
 \be
 H_0^{-+} = \sum_{x=1}^{b-1} h_{x x+1}^{-+} .
 \ee
This will be used in Section 3.

It is useful to introduce another parameter, $0<q<1$, such that
$q+q^{-1}=2\D$. The Hamiltonians defined in (\ref{H_0 kink}) and
(\ref{H_0 antikink}) in the spin 1/2 case, commute with a
representation of $SU_q(2)$. For $j>1/2$, the only obvious
conserved quantity is the total $S^3$-component, which commutes
with all the Hamiltonians defined above.

In the following we first deal with the kink Hamiltonian,
$H_0^{+-}$. There is a unique ground state for each value (sector)
$m$ of $S^3$, all of which have the same energy 0. The
eigenvalues, $m_x$, of $S_x^3$ are in  $\{-j,-j+1,\ldots,j-1,j\}$,
such that the total $S^3$-component takes the values
$m=\sum_{x\in[1,b]} m_x$. The eigenvectors of $S^3$ are denoted by
$|(m_x)\rangle$, and we have $S_y^3|(m_x)\rangle =
m_y|(m_x)\rangle$. Further, let
 $$
 w_m=\sqrt{2j\choose m+j}.
 $$
The unique ground states in the respective $S^3$-sectors are
called kink states which were found by Alcaraz, Salinas and
Wreszinski~\cite{ASW}. They are given by
 \be
 \psi_m = \sum_{\{m_x\}} \prod_{x=1}^b q^{-x(j-m_x)}w_{m_x}|(m_x)\rangle ,
 \ee
As the $\psi_m$ are not normalized, we also define
$\phi_m=\psi_m/\|\psi_m\|$. The sum over $\{m_x\}$ is restricted
to combinations such that $\sum m_x = m $. If $m=\pm jb$ is
maximal/minimal, then the state $\phi_m$ is the all spin-up/down
state, i.e. the magnetization profile in the $z$-direction,
$\langle\phi_m |S_x^3|\phi_m\rangle = \pm j$.
Both from a physical and mathematical point of view, the infinite
chain limit is the most interesting case. Clearly, more care has
to be taken when using an infinite volume Hamiltonian. The natural
tool is the {\sf GNS} representation. In our case here, there are
four (unitarily) inequivalent representations on Hilbert spaces,
which are also called {\it sectors}. These are the {\it (anti)kink
sectors} containing the infinite volume (anti)kink ground states,
and the {\it (anti)droplet sectors} with the all spin (down)up
ground state. As mentioned above, for the infinite volume limit it
is sufficient to study the effect of the perturbation for the
finite chain Hamiltonians, $H_0^{++}$ and $H_0^{+-}$. For more
details, we refer to \cite{KN1}.

For our purposes it will be very convenient to define the states
 \be
 \psi(z) = C \sum_{|m|\le jb} z^{jb-m} \psi_{m}.
 \ee
where $C$ is a normalization constant. They are product states,
i.e.
 \be
 \label{product}\psi(z) = \bigotimes_{x=1}^b \chi_x(z),
 \ee
with
 \be \label{chi}
 \chi_x(z) = (1+|z|^2q^{-2x})^{-j}
 \sum_{m_x=-j}^j (zq^{-x})^{j-m_x} w_{m_x} |(m_x)\rangle .
 \ee

The same construction can be carried through for the antikink
Hamiltonian. We denote the corresponding states for the antikink
Hamiltonian by $\tilde{\psi}(z)$. If we let
 \bea\lefteqn{\tilde{\chi}_x(z)} \label{anti}
 \\
 &=&(1+|z|^2q^{-2x})^{-j} \sum_{m_x=-j}^j (zq^{-x})^{j-m_x}  w_{m_x}
    |(-m_x)\rangle ,\nonumber
 \eea
then $\tilde{\psi}(z) = \bigotimes_{x=1}^b \tilde{\chi}_x(z)$ are
ground states of $H^{-+}$.

Now, let $\vec{B}=(B_1,B_2,B_3)$ be a magnetic field vector with
(real) parameters, and $V=\vec{B}\cdot \vec{S} = B_1 S^1 + B_2 S^2
+B_3 S^3$. Then the eigenvalues of $V$ are $\|\vec{B}\|\cdot m$
with $m= -j,-j+1,\ldots,j$. Define
 \bea H^{+-}(\vec{B})&=&H_0^{+-} + \vec{B}\cdot \vec{S}_y ,
 \\
 H^{++}(\vec{B})&=&H_0^{++} + \vec{B}\cdot \vec{S}_y .
 \eea
In the study of the spectrum of $H^{+-}(\vec{B})$ and
$H^{-+}(\vec{B})$, it is important to distinguish two cases:
$B_1^2+B_2^2>0$ and $B_1^2+B_2^2=0$. The ground state in these two
cases is described in Propositions \ref{prop:kink-1} and
\ref{prop:kink-2}, and the Remarks \ref{rem:1}.

\begin{prop}[Kink sector, $B_1^2+B_2^2>0$] \label{prop:kink-1}
Let $1\le y\le b$.
Then, the ground state of $H^{+-}(\vec{B})$ is the state $\psi(z)$
of eq (\ref{product}) with
$z=-\frac{\|\vec{B}\|+B_3}{B_1-iB_2}\,q^{y}$. Its energy is
$-j\|\vec{B}\|$.
\end{prop}
The proof is a combination of previously known results and a
straightforward calculation. See Appendix A.

\begin{figure}[t]
\resizebox{3.5in}{3in}{\includegraphics{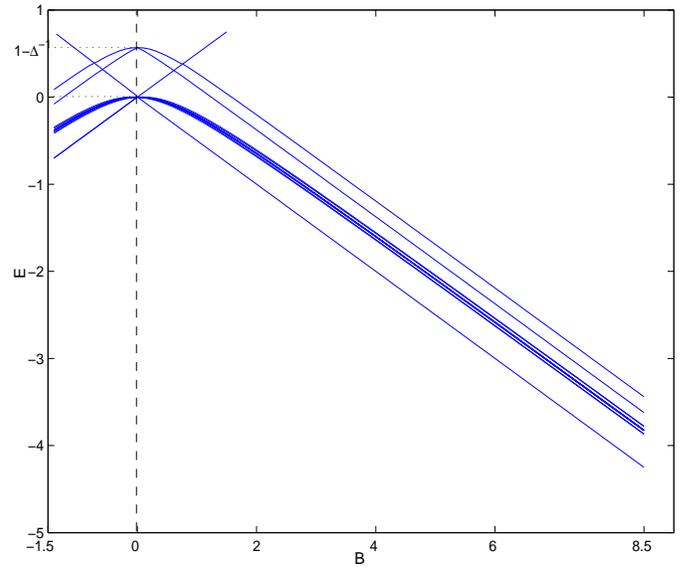}}
\caption{\label{fig2} Energy of the lowest 16 eigenvalues for the
spin $1/2$ kink Hamiltonian for the field $(B,0,0)$ on a chain of
13 sites and $\D=2.25$. Notice the ground state energy (straight
line downwards from 0), the gap to the 2nd eigenvalue, the energy
$\frac{1}{2}|B|$, and the energy $1-\D^{-1}-|B|/2$ of $\psi_e$. }
\end{figure}

\begin{figure}[t]
\resizebox{3.5in}{3in}{\includegraphics{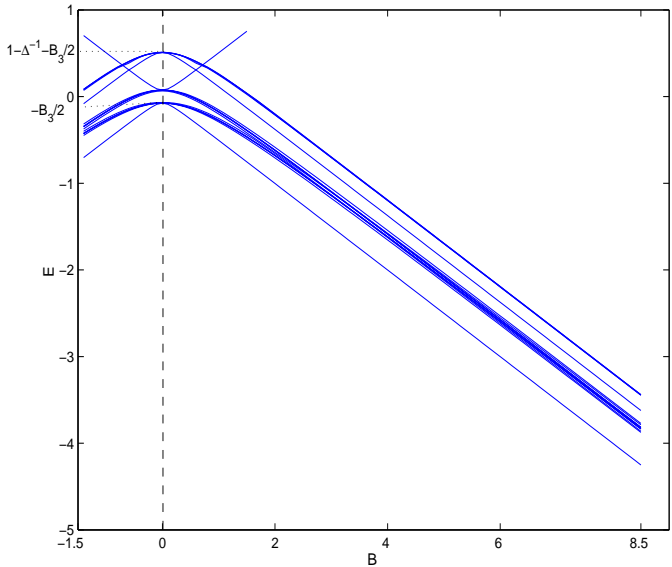}}
\caption{\label{fig2.5} Energy of the lowest 16 eigenvalues for
the spin $1/2$ kink Hamiltonian for the field $(B,0,A/6)$ on a
chain of 13 sites and $\D=2.25$; the $z$-component, $B_3=A/6=
\sqrt{1-\D^{-1}}/6$ is chosen small compared to the gap
$1-\D^{-1}$ of $H_0^{+-}$. Notice the ground state energy, the gap
above it, and the branches descending from $\pm B_3/2$, and
$1-\D^{-1}-B_3/2$. }
\end{figure}

\begin{figure}[t]
\resizebox{3.5in}{3in}{\includegraphics{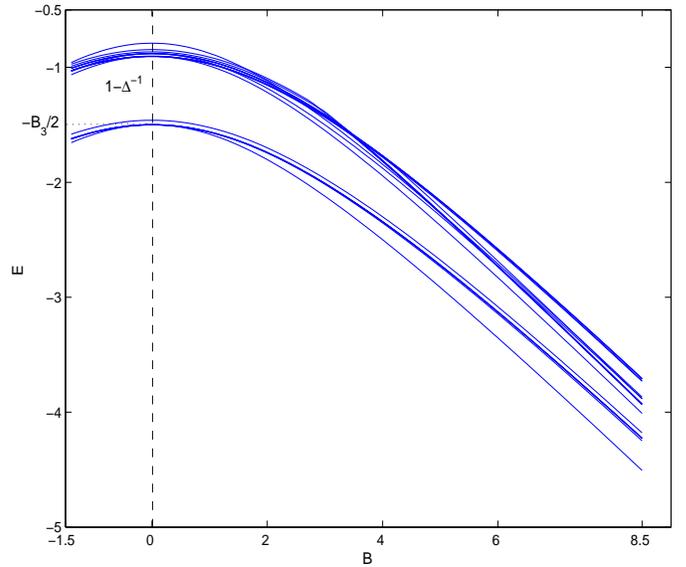}}
\caption{\label{fig3} Energy of the lowest 20 eigenvalues for the
spin $1/2$ kink Hamiltonian with the field $(B,0,3)$ on a chain of
13 sites and $\D=2.25$; the field in the $z$-direction is chosen
large compared to $1-\D^{-1}$. Notice that the branches stemming
from $B_3/2$ are too high in energy to be plotted here. We see the
ground state energy, the gap above it, and the energy of $\psi_e$
descending from $1-\D^{-1}-B_3/2$. In between there are $b-y+1=6$
states bending downwards from $-B_3/2=-1.5$. }
\end{figure}

\begin{remarks}\label{rem:1}
\begin{enumerate}
\setcounter{enumi}{-1}
\item The ground state of $H^{-+}(\vec{B})$ is the state $\tilde{\psi}(\tilde{z})$
of equation (\ref{anti}) with $\tilde{z}=-\frac{\|\vec{B}\|-B_3}{B_1+iB_2}\,q^{y}$.
This follows by a rotation by the angle $\pi$ with respect to the $x$-axis.

\item For simplicity, let us assume that $B_1^2+B_2^2=1$. If
$B_3=0$, then the ground state, $\psi(z=-(B_1+iB_2)\,q^{y})$, is a
kink state (exponentially) localized at the magnetic field at $y$;
among the spanning set of ground states $\psi(z)$, the
perturbation picks the one which is most localized at $y$. If
$B_3\neq0$, the extra term $B_3 S_y^3$ has the effect of shifting
the kink from $y$ by the distance $|\log_q{(\sqrt{1+B_3^2}+B_3)}|$
to the left if $B_3>0$, and to the right (by the same distance) if
$B_3<0$.

\item The proof also shows that the state $\psi(z)$ with
$z=+\frac{\|\vec{B}\| + B_3}{B_1-iB_2}\,q^{y}$ is an eigenstate
of $H^{+-}(\vec{B})$ with energy $+j\|\vec{B}\|$. We see this (it
is obviously linear if the field is in the $x$ or $y$ direction
only, see Figure \ref{fig2}) branch ascending from $|B_3|/2$ in
Figure \ref{fig2}-\ref{fig2.5}. In Figure \ref{fig3} it has too
high an energy to be among the plotted lowest eigenvalues.

\item Of special interest is the second-lowest eigenvalue, in
particular, whether there is a gap uniformly in the number of
sites, $b$, and how it depends on $\vec{B}$ and the anisotropy
$\D$. This will be treated in Section 3. We can extend a method
\cite{Na1} which was first applied to prove a gap for $H^{+-}_0$.

\item We discuss now qualitatively the low energy spectrum, and
assume for simplicity that $j=1/2$ and $B_2=0$. It was proven in
\cite{KN1} that the gap above the $b+1$ ground states of the
unperturbed Hamiltonian, $H_0^{+-}$, is equal to
$1-\cos{(\pi/b)}\D^{-1}$, which tends to $1-\D^{-1}$ in the
infinite chain limit.

If $B_3=0$, then there are $b$ eigenvalues (recall, $b$ is the
length of the chain) of $H^{+-}(B,0,0)$ descending from 0. This
can be seen as follows. We introduce the function
$N(B)=\chi_{(-\infty,0]}\left(H^{+-}(B,0,0) - \frac{1}{2}B{\bf
1}\right)$ counting the number of non-positive eigenvalues; here
$\chi_{(-\infty,0]}$ is the characteristic function of
$(-\infty,0]$. $N(B)$ is monotonically increasing, and equal to
$b+1$ for $|B| < 1-\D^{-1} + {\cal O}(b^{-1})$; this is guaranteed
by the gap above the ground states. In item \ref{item5} below we
calculate the lowest energy state, $\psi_e$, descending from
$1-\D^{-1}$ at $B=0$ with energy equal to
$1-\D^{-1}-\frac{1}{2}|B|$ (up to ${\cal O}(b^{-1})$). This state
is `parallel' to the ground state energy and intersects with the
state in item 2 at $|B|=1-\cos{(\pi/b)}\D^{-1}$, see Figure
\ref{fig2}.

We can say more about the average of these lowest $b+1$
eigenvalues by recalling the Min-max Principle, namely that their
average is a concave function in $B$. By symmetry  ($B\to-B$) its
(left and right) derivative is always negative and less than
$1/2$. Since the average is 0 at $B=0$ it continues to be
negative. In the infinite volume limit there is an infinite number
of ground states of $H_0^{+-}$, and eigenvalues for
$H^{+-}(B,0,0)$ have to accumulate at some value between
$-\frac{1}{2}|B|$ and $1-\D^{-1}-\frac{1}{2}|B|$, see Figure
\ref{fig2}, and in case $B_3\not=0$, see Figures
\ref{fig2.5}-\ref{fig3}.

\item If $B_3\not=0$, then there appear to be $b-y+1$ eigenvalues
descending from $-|B_3|/2$; $b-y+1$ is the number of kinks to the
left of $y$, see Figure \ref{fig2.5}. The next eigenvalues depend
on $B_3$. If $|B_3|<1-\D^{-1}$, then the next lowest $y$
eigenvalues descend from $-|B_3|/2$. If $|B_3|>1-\D^{-1}$, then
the state $\psi_e$ appears to be the next lowest in energy, see
Figure \ref{fig3}.

\item \label{item5} We can calculate the state mentioned in the
previous two items. Let $\psi_e=\sum_{x=1}^b a_x
S_x^-|\!\!\uparrow\rangle$ be the first excited state of
$H_0^{+-}$ in the one-overturned spin-sector. The coefficients
$a_x$ are a solution to the discrete Laplace equation (see
\cite{KN3}, or cf.~Appendix C by setting $B=0$). The energy of
$\psi_e$ is equal to the gap $1-\D^{-1}+{\cal O}(b^{-1})$. Now,
define
 $$
 \psi_e(z)=\sum_{n=0}^{b} z^n (S_q^- )^n \psi_e =\sum_{x=1}^b a_x
 S_x^- \psi(z).
 $$
By choosing $z$ as in Proposition \ref{prop:kink-1}, we obtain the
equation
 $$
 H^{+-}(\vec{B})\psi_e(z)=\left(1-\D^{-1}
 -\mfr{1}{2}\|\vec{B}\|\right)\psi_e(z) +{\cal
 O}\left(b^{-1}\right).
 $$
In Figure \ref{fig2}, this is the straight line parallel to the
ground state energy.
\end{enumerate}
\end{remarks}

\begin{figure}[t]
\resizebox{3.5in}{3in}{\includegraphics{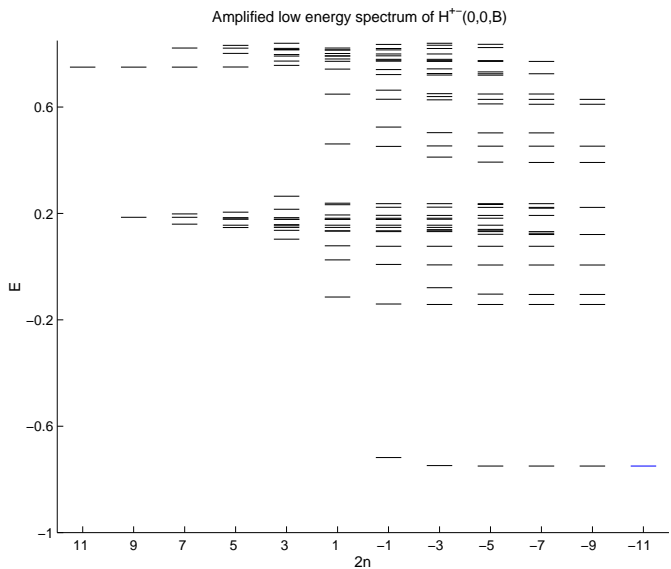}}
\caption{\label{fig4} Low energy spectrum in the spin $1/2$ kink
sector for the field $(0,0,1.5)$ on a chain of 11 sites with
$\D=2.25$. The high energy spectrum is similar to Fig.~\ref{fig6}
and will not be reprinted. $n$ is the total $z$-component. }
\end{figure}

Next, we consider the kink sector and magnetic fields of the form
$B_1=B_2=0,B=B_3\not=0$. Let $B>0$. Then, as $b$, the size of the
system increases, the ground state tends to the all spin-down
state, $|\!\downarrow\rangle$. This vector is no longer in the
infinite volume kink-sector very much as $e^{ikx}$ is not a
genuine (i.e., normalizable) eigenvector of the Laplacian on the
real line. In other words, $|\!\downarrow\rangle$ is part of the
continuous spectrum. So let us consider the orthogonal sequence of
kink states, $\psi_n$; $n$, as usual, is the total $z$-component.
Then, the sequence $\langle\psi_n,H^{+-}(0,0,B),\psi_n\rangle$
converges to $-jB$ as $n\to-\infty$. Since the spectrum is closed
and $-jB$ is the least possible eigenvalue it has to be the ground
state energy. Therefore, in the infinite chain limit, $-jB$ is
contained in the continuous spectrum, and is hence non-isolated.
We conjecture that there is no other continuous spectrum close to
$-j|B|$, and thus $-j|B|$ is purely an accumulation point of
eigenvectors. We do not give a proof of this here. Similarily, if
$B<0$, then the bottom of the spectrum is $jB$ and there is no gap
above the ground state, which is obviously the all spin-up state.
We illustrate the low energy spectrum in Figure \ref{fig4}.

Let us collect our results in the following proposition:
\begin{prop}[Kink sector, $B_1^2+B_2^2=0$]\label{prop:kink-2}
The bottom of the spectrum of $H^{+-}(0,0,B)$ is equal to $-j|B|$,
which is part of the continuous spectrum. Excitations above the
ground state are gapless.
\end{prop}
Now we consider the Hamiltonian $H^{++}(\vec{B})$. It is useful to
decompose this as a sum of a kink and anti-kink Hamiltonian
($b\ge3$):
 \beax H^{++}_{[1,b]} + 2jA\left(-j{\bf 1} + S_y^3\right)
 = H^{+-}_{[1,y]} +
    H^{-+}_{[y,b]}  ,
 \eeax
and thus
 \bea \label{decomp} H_{[1,b]}^{++}(\vec{B}) - 2j^2 A {\bf 1}
&=&H^{+-}_{[1,y]}(\mfr{B_1}{2},\mfr{B_2}{2},\mfr{B_3}{2}-jA) \nonumber
\\
&&+ H^{-+}_{[y,b]}(\mfr{B_1}{2},\mfr{B_2}{2},\mfr{B_3}{2}-jA).
 \eea
We start with the case $B_1^2+B_2^2>0$. As in the kink sector we
will find a unique ground state. Let
 $$
 z=-\frac{\|(B_1,B_2,B_3-2jA)\| + B_3 -2jA}{B_1-iB_2}q^y,
 $$
then according to Proposition \ref{prop:kink-1}, $\psi_{[1,y]}(z)$
is the unique ground state of
$H^{+-}_{[1,y]}(B_1/2,B_2/2,B_2/2-jA)$, while the anti-kink state
$\tilde{\psi}_{[y,b]}(\tilde{z})$ is the corresponding ground state of
$H^{-+}_{[y,b]}(B_1/2,B_2/2,B_3/2-jA)$. They are both product
states which happen to satisfy
$\chi_y(z)=(-1)^{2j}\tilde{\chi}_y(\tilde{z})$, because
 \beax\lefteqn{\mfr{1}{2}(B_1 S_y^1 + B_2 S_y^2 + (B_3 - 2jA)
S_y^3)\chi^\sharp(z^\sharp) }
\\
&=&-\mfr{j}{2}\|(B_1, B_2, B_3 -2j A)\| \,\chi^\sharp(z^\sharp),
 \eeax
where $\chi^\sharp(z^\sharp)$ stands for either $\chi_y(z)$ or $\tilde{\chi}_y
(\tilde{z})$. Thus
 $$
 \psi(\vec{B})=\bigotimes_{x=1}^y \chi_x(z) \otimes \bigotimes_{x=y+1}^b
 \tilde{\chi}_x(\tilde{z})
 $$
is the unique ground state of $H^{++}_0(\vec{B})$ with energy
$-j\|(B_1,B_2,B_3-2j A)\| + 2j^2 A$.

Similar to the kink-sector, the vector
 $$
 \bigotimes_{x=1}^y \chi_x(-z) \otimes \bigotimes_{x=y+1}^b
 \tilde{\chi}_x(-\tilde{z})
 $$
is another eigenstate with energy ${j}\|(B_1,B_2,B_3-2j A)\| + 2j^2
A$.

Finally, we come to the case $B_1=B_2=0,B=B_3$. When $B=0$, it was
proven in \cite{NS} that for spin $1/2$, and in the infinite
volume limit $1-\D^{-1}$ is the gap above the ground state. It can
also be shown that there exists a gap, $\d$, for higher spins
although no precise estimates are known. This implies that
uniformly in the size of the lattice, that the all spin-up vector
is the unique ground state for $B < B_c = \d/(2j)$, where $\d$ is
the (strictly positive) gap of $H^{++}_0$.
As we mentioned in the Introduction, the value $B=A/(2j)$ is very
particular and interesting. It has been analyzed \cite{St2} in the
context of droplet states for spin $1/2$ but again the method
extends to general $j$. In fact, the set of ground states is
infinitely degenerate (in the infinite volume) and consists of
pairs of symmetric kink-antikink states (i.e.~droplets), all of
which have the same energy $j A$. The magnetization profile in the
$z$-direction is symmetric with respect to the center of the field
at $y$. Excitations are gapless because large droplets which are
antisymmetric with respect to $y$ come arbitrarily close in energy
to $j A$.

Since the ground state energy is concave, and since for $B=0$ and
$B=A/2j$, the all spin-up vector is a ground state, we conclude
that for all $B < A/(2j)$, the all spin-up vector is the unique
ground state. Numerical experiments for spin $1/2$ indicate that
in the region $B < A$ the eigenvalues are ordered by their total
$S^3$-value such that the second-lowest eigenstate is in the
sector with one overturned spin, and has energy ${\cal E}_-(B)=1-
\sqrt{\D^{-2}+B^2} + \frac{1}{2}|B|$, see (\ref{excit}); its
(infinite volume) derivation is given in Appendix C. The
third-lowest eigenvector is in the sector with two overturned
spins, and so on. The lowest eigenvalues with respect to the total
$S^3$-component accumulate at the line $A - B/2$. They all meet at
the critical value $A$, where the ground state becomes infinitely
degenerate. Assuming that this ordering holds true, we conjecture
that the gap for spin $1/2$ and $B\le A$ equals $1-\sqrt{\D^{-2} +
B^2} +\frac{1}{2}(|B|-B)$, which converges to $1-\frac{1}{2\D}$
for large $-B$, and vanishes at $B=A$.

For $B > A/(2j)$, the all spin-down state is the unique ground
state with energy $A/(2j) - jB $, which is part of the continuous
spectrum; in fact, $A/(2j) - jB $ is purely an accumulation point
of eigenvectors. It seems that for $B>A/2j$ the eigenvalues are
also ordered according to their total $z$-component, $n$, but this
time in the opposite way. I.e.~lower $n$ means lower energy, and
clearly the lowest is the all spin-down state. Similar to the kink
sector, we will not prove here that the rest of the continuous
spectrum is separated from the ground state.

The situation is illustrated in Fig.~\ref{fig5}-\ref{fig7} for
$j=1/2$, cf.~also Figure \ref{fig1}. Let us summarize our results
in the following proposition:

\begin{figure}[t]
\resizebox{3.5in}{3in}{\includegraphics{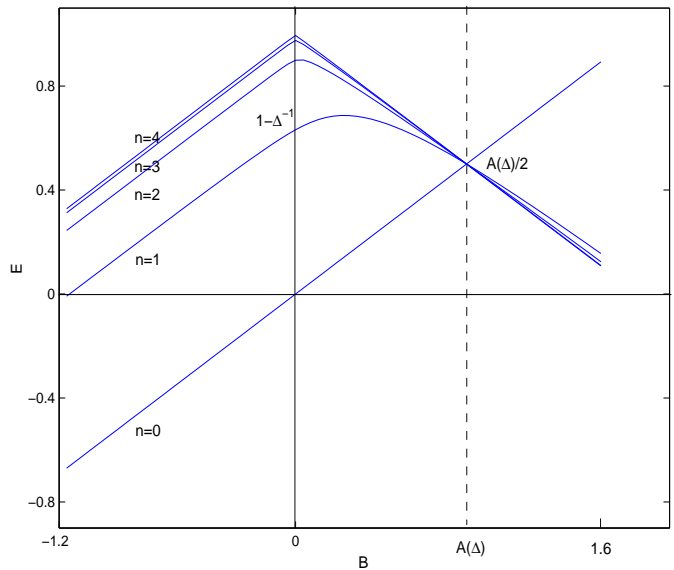}}
\caption{\label{fig5} The five lowest eigenvalues in the spin
$1/2$ droplet sector for a chain of 13 sites with magnetic field
$(0,0,B)$ and $\D=2.25$. The eigenvalues are numbered by $n$, the
number of overturned spins; e.g.~$n=0$ means the all spin-up
state. }
\end{figure}



\begin{figure}[t]
\resizebox{3.5in}{3in}{\includegraphics{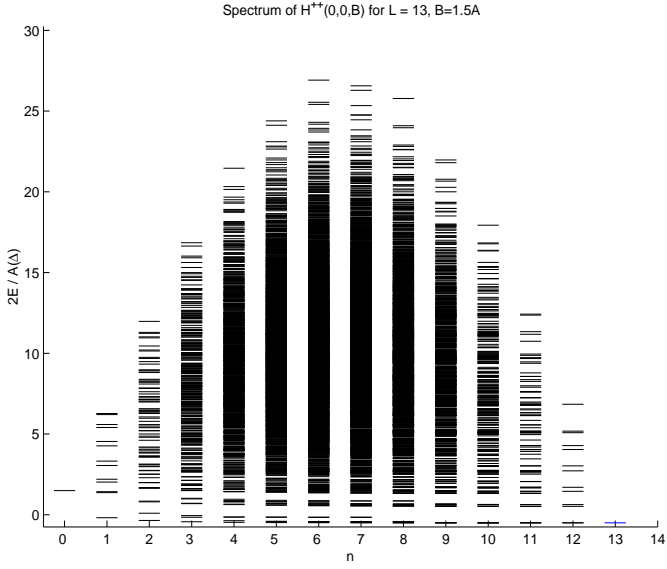}}
\caption{\label{fig6}Here, we plot the full spectrum of the spin
1/2 droplet Hamiltonian $H^{++}(0,0,1.5A)$ for 13 sites, and
$\D=2.25$. The index $n$ is the number of overturned spins. The
ground state is the all spin-down state, i.e.~$n=13$.}
\end{figure}

\begin{figure}[t]
\resizebox{3.5in}{3in}{\includegraphics{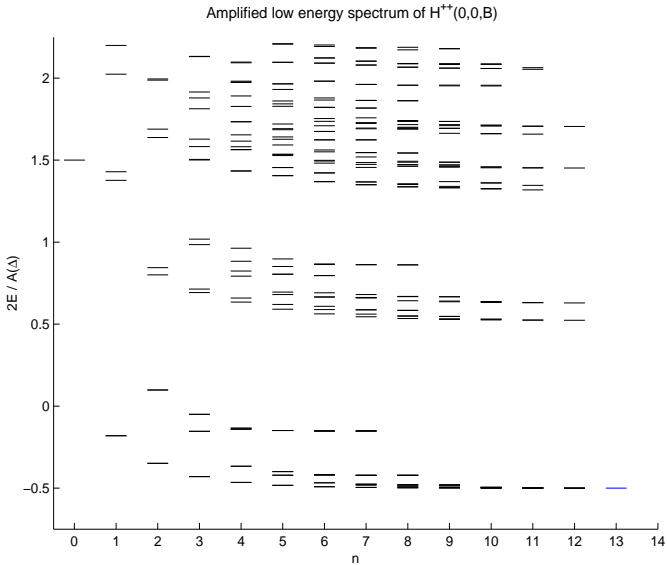}}
\caption{\label{fig7} Here, we amplify the low energy spectrum for
the same Hamiltonian as in Figure \ref{fig6}. At $n=13$ we have
indicated the ground state. Clearly, one sees monotonicity of
energy vs $n$.}
\end{figure}

\begin{prop}[Droplet sector] \label{droplet sector}
The ground state of the droplet Hamiltonian,
$H^{++}(\vec{B})$, on a chain of length $b\ge3$ depends on the
magnetic field $\vec{B}$ in the following way:
\begin{enumerate}

\item If $B_1^2+B_2^2>0$, then the ground state is unique. The ground
state energy is $-{j}\|(B_1,B_2,B_3 - 2jA)\| + 2j^2 A$.

\item If $B_1=B_2=0,B=B_3$, and $B_c = A/(2j)$, then for $B < B_c$ the unique
ground state of $H^{++}(0,0,B)$ is the all spin-up vector with
energy $jB$. For $B = B_c$ the ground states are droplet states
which are (in the thermodynamic limit) infinitely degenerate with
energy $jA$. In infinite volume, excitations above these ground
states are gapless. For $B > B_c$, the all spin-down state is
the unique ground state, which is an accumulation point of
eigenvectors. Hence, excitations are gapless.
\end{enumerate}
\end{prop}

\section{Estimate for the spectral gap in the case $j=1/2$}

Here we prove a uniform lower bound on the difference between the
ground state energy and the energy of the first excited state for
the spin $1/2$ Hamiltonians $H^{+-}_{[1,b]}(\vec{B})$ and
$H^{++}_{[1,b]}(\vec{B})$ on a finite chain $[1,b]$ with the
impurity field at $y$.

Before we prove these gap inequalities we introduce the methods
which were invented in \cite{Na1} and \cite{Na2}. Let $\C_i,
i=0,\ldots,N$ be a sequence of connected intervals with
$\bigcup_{i=0}^N \C_i = [1,b]$, and such that two intervals have
at most one lattice point in common. Let $h_{\C_i}\ge 0$ be some
(local) Hamiltonians acting on $\mathbb C^{2|\C_i|}$, and define
\be H_{[1,b]} = \sum_{i=0}^N h_{\C_i}. \ee $H_{[1,b]}$ acts on
${\cal H}_b = \bigotimes_{i=1}^b \mathbb C^2$. We assume that
$\ker{H_{[1,b]}}\not=\{0\}$. Let $\g_i$ denote the gap of
$h_{\C_i}$, i.e.~the smallest non-zero eigenvalue of $h_{\C_i}$.
It is clear that \be \ker H_{[1,b]} = \bigcap_{i=0}^N \ker
h_{\C_i} . \ee Let $\L\subset[1,b]$, then we define $G_\L$ to be
the orthogonal projection onto the \be \ker \sum_{i:\C_i\subset
\L} h_{\C_i}. \ee We use the convention that if
$\C_i\not\subset\L$ for any $i$, then we set $G_\L=\bf 1$. From
these definitions we derive the following properties:
\begin{enumerate}
\item $G_{\L} G_{\L'} = G_{\L'} G_{\L} = G_{\L'}$ if $\L\subset\L'$.
\item $G_{\L} G_{\L'} = G_{\L'} G_{\L}$ if $\L\cap\L'=\emptyset$.
\item $h_{\C_i}\ge \g_i ({\bf 1} - G_{\C_i})$.
\end{enumerate}
Next we define the intervals $\L_i=\bigcup_{j\le i}\C_j$, and operators
$E_i, i\ge0$, on ${\cal H}_b$ by
 \be E_i = \left\{\begin{array}{lcl}{\bf 1} - G_{\L_0}& \mbox{for} &i=0
                   \\G_{\L_i} - G_{\L_{i+1}}&\mbox{for} &1\le i<N
           \\G_{[1,b]}&\mbox{for}&i=N
         \end{array} \right. .
 \ee
These operators are mutually commuting projections adding up to ${\bf 1}$, i.e.
 \be
 E_i^* = E_i,\quad E_iE_j=\d_{ij} E_i,\quad \sum_{i=0}^N E_i ={\bf 1}.
 \ee
The key assumption in order to deduce a gap for $H_{[1,b]}$ from
the gaps of $h_{\C_i}$ is the following assumption:
\begin{assumption} \label{assumption} There exists a positive constant
$\e$ such
that $0\le \e < 1/\sqrt{2}$ and
 \be \label{ass1}
 E_i G_{\C_{i+1}} E_i \le \e^2 E_i, \quad 0\le i\le N-1 ,
 \ee
or equivalently,
 \be \label{key est}
 \|G_{\C_{i+1}}E_i\|\le \e, \quad 0\le i\le N-1.
 \ee
Further, we assume that the gaps, $\g_i$, for the local
Hamiltonians are bounded from below, i.e.~$\g_i\ge\g>0$.
\end{assumption}

The conditions (\ref{ass1}) and (\ref{key est}) are equivalent due
to $G_{\C_{i+1}}E_i =G_{\C_{i+1}}G_{\L_i}$ $-G_{\L_{i+1}}$.

Now we are ready to state the main theorem which we apply in all
three case below.

\begin{thm}[Nachtergaele \cite{Na1}] With the above definitions
and under the assumptions in (\ref{key est}) let $\psi$ be orthogonal to 
the ground states of $H_{[1,b]}$. Then
 \be\label{general gap}
 (\psi, H_{[1,b]}\psi) \ge \g (1-\sqrt{2}\e)^2\|\psi\|^2.
 \ee
I.e.~the gap in the spectrum of $H_{[1,b]}$ above 0 is at least
$\g (1-\sqrt{2}\e)^2$.
\end{thm}

\begin{proof}Let $\psi$ be orthogonal to the ground state, i.e.
$G_{[1,b]}\psi=0$. Then,
$\|\psi\|^2 = \sum_{0\le n< N}\|E_n \psi\|^2$.

We estimate $\|E_n \psi\|^2$ in terms of $(\psi,H_{\C_n}(\vec{B})\psi)$
as follows. First notice, that for $m\le n-2$, or $m\ge n+1$,
$E_m G_{\C_n} = G_{\C_n} E_m$.
Now we insert $G_{\C_n}$ and the resolution of identity, $\{E_n\}$, and we get
 \bea
 \lefteqn{\|E_n \psi\|^2}
 \label{3.11}\nonumber\\
 &=& (\psi,({\bf 1}- G_{\C_n}) E_n\psi) + (\psi,\sum_{0\le m< N}
 E_m G_{\C_n} E_n
                         \psi)
 \nonumber\\
 &=&(\psi,({\bf 1}- G_{\C_n}) E_n\psi) + ((E_{n-1} +E_{n})
 \psi,G_{\C_n} E_n\psi).\hspace{.7cm}
 \eea
Let $c_1,c_2>0$, then
 \bea \lefteqn{\|E_n \psi\|^2}
 \nonumber
 \\ &\le&\frac{1}{2c_1} (\psi, ({\bf 1} - G_{\C_n})\psi)
      + \frac{c_1}{2} (\psi,E_n\psi)
 \label{3.12}
 \\
 &&+\frac{1}{2c_2} (\psi,E_n G_{\C_n}E_n\psi) + \frac{c_2}{2}
 (\psi,(E_{n-1} + E_{n})^2\psi), \nonumber
 \eea
where we used the inequality
 $$
 |(\phi_1,\phi_2)| \le \frac{1}{2c} \|\phi_1\|^2 + \frac{c}{2} \|\phi_2\|^2
 $$
in both terms of (\ref{3.11}). Let $\gamma = \min{\{\g_{\C_i}\}}$ be the
minimum
of the gaps of the Hamiltonians $h_{\C_i}$.

The first term on the rhs of (\ref{3.12}) is less than
$\frac{1}{2c_1 \gamma} (\psi, h_{\C_n} \psi)$. Now, assuming the key
estimate,
$\|G_{\C_n} E_n\| < \e$, we see that
 \beax\lefteqn{\left(2-c_1 - \frac{\e^2}{c_2}\right) \|E_n\psi\|^2 - c_2
 \|(E_{n-1}+E_{n})\psi\|^2}
 \\
 &\le&\frac{1}{c_1\gamma} (\psi,h_{\C_n} \psi) .\hspace{4cm}
 \eeax
We sum over $n$ using $\|\psi\|^2 = \sum_{0\le n<N}\|E_n \psi\|^2$ from
above and get
 $$
 \left(2-c_1 - \frac{\e^2}{c_2} - 2c_2\right) \|\psi\|^2 - c_2 \|\psi\|^2 \le
 \frac{1}{c_1\gamma} (\psi, H_{[1,b]}\psi) .
 $$
Finally, we optimize the constants $c_1,c_2$ yielding
$c_1=1-\e\sqrt{2}, c_2=\e/\sqrt{2}$. This proves the gap
inequality.
\end{proof}

In all the upcoming proofs on the various gaps we use the same
definition of subsets $\C_n,\L_n$ of $[1,b]$ and projections $G_n,
E_n$. As usual $y\in[1,b]$ denotes the spot of the magnetic field.
Let $n_l,n_r$ be some non-negative integers such that
$n_r+n_l\ge1$, and assume that $n_r>0$; the choice of $n_l,n_r$ in
general will depend on $\D$ and $\vec{B}$. The idea behind the
definition of $\C_n$ is that we cover the chain $[1,b]$ by adding
points to an initially chosen interval $\C_0=[y-n_l,y+n_r]$ in an
alternating manner. First we add a point to the right of $\C_0$,
then to the left until we reach the point 1. Then we add points
only to the right of $\C_{2(y-n_l-1)}$ until we finish at $b$.
More precisely, we define the sets $C_n$ in the following way:

\begin{defin} \label{def}
Let $\C_0=[y-n_l,y+n_r]$, where we may assume that $y-n_l-1\le b-y-n_r$
such that
$\C_0\subset [1,b]$. The intervals for $n>0$ are then
$\C_1 = [y+n_r,y+n_r+1],\C_2 =
[y-n_l-1,y-n_l],\ldots, \C_{2(y-n_l-1)} = [1,2], \C_{2(y-n_l)-1} =
[2y+n_r-n_l-1,2y+n_r-n_l],\ldots,\C_{b-(n_l+n_r)-1} = [b-1,b]$.
\end{defin}

We start with the kink case.

\begin{prop}[Kink sector, $B_1^2+B_2^2>0$] \label{prop3.1}
Let $\psi$ be orthogonal to the ground state of the kink Hamiltonian,
$H^{+-}_{[1,b]}(\vec{B})$, on a chain of length $b$. Then, there exists a
strictly positive
function $g^{+-}(\vec{B},\D)$ and a function $0\le\e(\vec{B},\D)<1/\sqrt{2}$,
which are both independent of $b$, such that the following gap inequality is
satisfied
 \bea \label{gap alpha}
 (\psi,H^{+-}_{[1,b]}(\vec{B})\psi)
    \ge g^{+-}(\vec{B},\D) (1-\sqrt{2}\e(\vec{B},\D))^2\|\psi\|^2. \hspace{.7cm}
 \eea
\end{prop}
\begin{proof}
First we shift the ground state energy to be 0, and define the new Hamiltonian
 $$
 H_{[1,b]}(\vec{B}) = H^{+-}_{[1,b]}(\vec{B}) - \mfr{\|\vec{B}\|}{2} {\bf 1} .
 $$
With the set-up from Definition (\ref{def}) we
can write the Hamiltonian, $H_{[1,b]}(\vec{B})$, in the following form:
 \beax
 H_{[1,b]}(\vec{B})  &=& \sum_{i=0}^{b-(n_l+n_r)-1} h_{\C_i}
 \eeax
using
 \beax
  h_{\C_i} &=&\left\{\begin{array}{lcr}H_{\C_0}(\vec{B})&\mbox{ for }& i=0\\
  h_{x x+1}^{+-}&\mbox{ for }& i>0
               \end{array},\right.
 \eeax
where $\C_i = [x, x+1]$ for some $1\le x<y-n_l$ or $y+n_r \le x < b$, and with
$h_{x x+1}^{+-}$ from (\ref{h^{+-}}).

We can express the gap conditions as
\begin{enumerate}
\item $H_{\L_0}(\vec{B}) \ge g_{\L_0}^{+-}(\vec{B},\D) ({\bf 1} - G_{0})$,
where
$g_{\L_0}^{+-}(\vec{B},\D)$ is the gap for the finite chain Hamiltonian,
$H^{+-}_{\L_0}(\vec{B})$.
\item $h_{x x+1}^{+-} = {\bf 1} - G_{[x x+1]}$ for $1\le x<y-n_l$
and $y+n_r \le x< b$.
\end{enumerate}
Let $\g = \min{\{g_{\L_0}^{+-}(\vec{B},\D),1\}}$ which is strictly positive.
Finally, we need to verify the second condition in Assumption
\ref{assumption}, and define
 \be
 C_n:= \sup_{0\not=\psi\in{\cal H}_{\Lambda_{n+1}} : E_n\psi = \psi}
          \frac{\|G_{\C_n}\psi\|}{\|\psi\|}.
 \ee
So let $\psi$ satisfy
 \be
 G_n\psi = \psi\quad\mbox{and}\quad G_{n+1}\psi = 0 .
 \ee
First, let $n=2m, 0\le m\le y-n_l-1$; the case $n\ge 2(y-n_l)$ is similar, and
the case of odd
$1\le n < 2(y-n_l)$ will be considered later.

Let $\psi$ be a ground state of $\L_n$, i.e.~$G_n\psi=\psi$
such that $G_{n+1}\psi=0$. Then
with the definition from (\ref{chi})
 $$
 \psi = \bigotimes_{i={y-n_l-m}}^{y+n_r+m} \chi_i(z) \otimes
 \chi^\perp_{y-n_r+m+1}(z),
 $$
where $\chi^\perp_{x}(z)$ is perpendicular to $\chi_x(z)$.
Let us make some definitions and call
$f:=-\frac{\|\vec{B}\| +B_3}{B_1-iB_2}$,
$\chi:=\chi_{y+n_r+m}(z) = |\!\!\uparrow\rangle +
f q^{-n_r-m}|\!\!\downarrow\rangle$,
$\chi^\perp:=\chi_{y+n_r+m}(z) = f q^{-n_r-m-1}$ $|\!\!\uparrow\rangle -
|\!\!\downarrow\rangle$,
and $(1+q^2)^{1/2}|\xi\rangle =  q|\!\!\uparrow\!\downarrow\rangle -
|\!\!\downarrow\!
\uparrow\rangle$.
Then,
 \beax \frac{\|G_{\C_n}\psi\|^2}{\|\psi\|^2}
 &=& 1- \frac{\||\xi\rangle\langle\xi|\chi\otimes
      \chi^\perp\rangle\|^2}{\|\chi\otimes\chi^\perp\|^2}
 \\
 &=&1 - \frac{q^2}{1+q^2}
 \frac{1+|f|^2 q^{-2(n_r+m+1)}}{1+|f|^2 q^{-2(n_r+m)}}.
 \eeax
Let $m=0$, then we choose $n_r$ such that rhs is less than $1/2$.
The condition for $C_0<1/\sqrt{2}$ is thus $|f| q^{-n_r} > 1$. By
monotonicity it is clear that the condition,
$C_{n=2m}<1/\sqrt{2}$, holds true for $m\ge0$.

Now we come to odd integers, $n=2m+1$ with $0\le m\le y-n_l-1$.
Let $\psi$ satisfy $G_n \psi =\psi$, and $G_{n+1}\psi=0$, then
$\psi$ is of the form $\psi = \chi^\perp_{y-n_l-m-1}(z) \otimes
\bigotimes_{i=y-n_l-m}^{y+n_r+m} \chi_i(z)$. We have thus
 \beax
 \frac{\|G_{\C_n}\psi\|^2}{\|\psi\|^2} &=& 1- \frac{1}{1+q^2}
      \frac{1+|f|^2 q^{2(n_l+m+1)}}{1+|f|^2 q^{2(n_l+m)}} .
 \eeax
Let $m=0$, then we choose $n_l$ such that rhs is less than $1/2$.
This is accomplished if $1>|f| q^{n_l}$. By monotonicity,
$C_{n=2m+1}<1/\sqrt{2}$ for $m\ge0$. Our condition for the choice
of $n_l,n_r$ is thus
 $$
 q^{n_r} < \left|\frac{\|\vec{B}\|+B_3}{B_1-iB_2}\right|
<q^{-n_l}.
$$
\end{proof}

\begin{remark}
It is clear that there always exist integers $n_l$ and $n_r$ such
that $1> |f| q^{n_l}$ and $1<|f| q^{-n_r}$ are satisfied. Now
suppose that $q<\left|\frac{\|\vec{B}\|+B_3}{B_1-iB_2}\right|<1$,
then we may choose $n_l=0$ and $n_r=1$. In this case, 
$g^{+-}(\vec{B},\D)=g_2^{+-}(\vec{B},\D)$ is found explicitly in 
Appendix B, see (\ref{kinkgap}). If $B_3=0$, then one needs to choose
$n_l=n_r=1$ and diagonalize a three-site Hamiltonian which we will not 
do here.
\end{remark}

\begin{prop}[Droplet sector] $B_1^2+B_2^2>0$.
Let $\psi$ be orthogonal to the ground state of the droplet Hamiltonian,
$H^{++}_{[1,b]}(\vec{B})$, on a chain of length $b$. Then, there exists a
strictly positive
function $g^{++}(\vec{B},\D)$ and a
positive function $0\le\e(\vec{B},\D)<1/\sqrt{2}$,
which are both independent of $b$, such that the following gap inequality is
satisfied
 \be
 (\psi,H^{++}_{[1,b]}(\vec{B})\psi) \ge g^{++}(\vec{B},\D)
    (1-\sqrt{2}\e(\vec{B},\D))^2\|\psi\|^2.
 \ee
\end{prop}

\begin{proof}
First, we need to shift the ground state energy, and define a new Hamiltonian
 $$
 H_{[1,b]}(\vec{B})
 = H^{++}_{[1,b]}(\vec{B}) + \mfr{1}{4}\|(B_1,B_2,B_3 - A)\| -
                       \mfr{1}{4} A.
 $$
Using the sets from Definition (\ref{def}) we have the decomposition
 \beax
 H_{[1,b]}(\vec{B})&=&\sum_{i=0}^{b-(n_l+n_r)-1} h_{\C_i},
 \eeax
with
 \beax
 h_{\C_i}=\left\{\begin{array}{ll}H_{\C_0}(\vec{B})&\mbox{ for } i=0 \\
 h_{x x+1}^{+-}&\mbox{ for some $x$: } 1\le x<y-n_l\\
 h_{x x+1}^{-+}&\mbox{ for some $x$: } y+n_r \le x < b ,
 \end{array}\right.
 \eeax
depending on whether $\C_i$ is to left (right) of $y$. $h_{x
x+1}^{+-}$ and $h_{x x+1}^{-+}$ are taken from equations
(\ref{h^{+-}}), respectively (\ref{h^{++}}). We have the following
gap properties:
\begin{enumerate}
\item $H_{\L_0}(\vec{B}) \ge g^{++}_{\C_0}(\vec{B},\D) ({\bf 1} - G_0)$,
where $g^{++}_{\C_0}(\vec{B})$
is the gap for the Hamiltonian, $H_{\C_0}(\vec{B})$.
\item $h_{x x+1}^{+-} = {\bf 1} - G_{[x x+1]}$ for $1\le x<y-n_l$.
\item $h_{x x+1}^{-+} = {\bf 1} - G_{[x x+1]}$ for $y+n_r \le x< b$.
\end{enumerate}
We are left with verifying the key estimate (\ref{key est}). So
let $0\not=\psi$ satisfy $G_n\psi=\psi$ such that $G_{n+1}\psi=0$.
If the interval $C_{i+1}$ is to the left of $y$ then we have the
same situation as in the previous proof with the condition
$1>|f|q^{n_l}$ and the slightly modified 
$f=\frac{\|(B_1,B_2,B_3-A)\| +B_3-A}{B_1-iB_2}$.

If the interval $C_{i+1}$ is to the right of $y$, then we will
arrive at the same condition for $n_r$, namely $1>|f|q^{n_r}$. This 
is true by symmetry but one can easily derive this in the very same 
way we did in the other case.
\end{proof}

\begin{remark} The same remarks are in order here for the droplet
Hamiltonian. So let us suppose that
$\left|\frac{\|(B_1,B_2,B_3-A)\|+B_3-A}{B_1-iB_2}\right| < 1$,
or equivalently, $A\ge B_3$, then we choose $n_l=0$ and $n_r=1$. 
In this case, $g^{++}(\vec{B},\D) = g_2^{++}(\vec{B},\D)$ is explicitly
calculated in Appendix B, see (\ref{dropletgap1}). 
\end{remark}

\begin{prop}[Droplet sector] Let $B_1=B_2=0$, and $B<A$.
Let $\psi$ be orthogonal to the all spin-up ground state of the droplet
Hamiltonian,
$H^{++}_{[1,b]}(0,0,{B})$, on a chain of length $b\ge3$, and let
$g^{+}_3(B,\D)$ be the gap for the
three-site Hamiltonian from eq (\ref{dropletgap2}). Then,
 \be
 (\psi,H^{++}_{[1,b]}({B})\psi) \ge 2 g^{+}_3(B,\D) \left(\frac{1}{\sqrt{2}}-
 \sqrt{\frac{q}{q+q^{-1}}}\right)^2\|\psi\|^2.
 \ee
\end{prop}

\begin{proof}
Again, we need to shift the ground state energy, and define a new Hamiltonian
$$H_{[1,b]}(B) = H^{++}_{[1,b]}(0,0,B) - \mfr{B}{2} {\bf 1}.
$$
As before, we use the same decomposition of $H_{[1,b]}(B)$ into local
Hamiltonians,
$h_{\C_i}$.
The first gap condition of $\C_0$ has to be changed into
 $$
 H_{\L_0}(B) \ge g^{+}_{\C_0}(B,\D) ({\bf 1} - G_0).
 $$
We only need to compute
 $$
 C_n:= \sup_{0\not=\psi\in{\cal H}_{\L_{n+1}} : E_n\psi = \psi}
        \frac{\|G_{\C_n}\psi\|}{\|\psi\|}.
 $$
So let us take a (non-zero) vector $\psi$ such that $G_n\psi=\psi$
and $G_{n+1}\psi=0$. If $\C_{n+1}$ is to the left of $y$, then
$\psi = |\!\downarrow\rangle\otimes|\!\uparrow
\cdots\uparrow\rangle$, and $G_{\C_n} = {\bf 1} -
|\xi\rangle\langle\xi|$. Then
$$\frac{\|G_{\C_n} \psi\|^2}{\|\psi\|^2} = \frac{1}{1+q^2},
$$
which is less than $1/2$, and $(1-\sqrt{2}\e)^2= 2\left(\frac{1}{\sqrt{2}}-
\sqrt{\frac{q}{q+q^{-1}}}\right)^2$.

By symmetry this is also the condition if $\C_{n+1}$ is to the right of $y$.
More precisely,
$\psi=|\!\uparrow\cdots\uparrow\rangle \otimes |\!\downarrow\rangle$, and
$G_{\C_n} = {\bf 1} - |\zeta\rangle\langle\zeta|$
with $(1+q^2)^{{1}/{2}}|\zeta\rangle  =
q |\!\downarrow\uparrow\rangle - |\!\uparrow\downarrow\rangle$.

By choosing $\C_0=[y-1,y+1]$, we have verified the statement. The
three-site gap, $g^{+}_{3}(B,\D)$ is calculated in Appendix B.
\end{proof}

\begin{appendix}

\section{Proof of Proposition \ref{prop:kink-1}}

\begin{proof} First, it is clear that the bounded perturbation
$V=\vec{B}\cdot\vec{S}_y$ can shift the ground state energy of
$H_0^{+-}$ by no more than its norm, $j\|\vec{B}\|$. We claim, and
show below, that the product state
$\psi(z=-\frac{\|\vec{B}\|+B_3}{B_1-iB_2}\,q^{y})$, which is a
ground state of $H_0^{+-}$, is a also a ground state of $V$.
Therefore, we have found a ground state of $H_0^{+-}(\vec{B})$.
That it is the unique ground state follows by combining two facts:
1) $\psi(z)$ is the unique kink state with this property, which we
will show, and 2) the vectors $\psi(z)$, for arbitrary complex
$z$, span the full ground state space of $H_0^{+-}$ \cite{GW,KN3}
and there is gap to the rest of the spectrum \cite{KN1,KNS}. So,
it only remains to prove that among all vectors $\psi(z)$, there
is a unique one that is a ground state and that the corresponding
value of $z$ is as stated in the proposition.

Since $\psi(z)$ is of product form, and $V$ acts non-trivially
only at site $y$, we are left to show that
 \be \label{7}
 \vec{B}\cdot\vec{S}_y \chi_y(z) = -j\|\vec{B}\|\chi_y(z) .
 \ee
Without loss of generality, we may assume that $\|\vec{B}\|=1$.
Then, $zq^{-y}=-\frac{1+B_3}{B_1-iB_2}$. Now, checking all $2j+1$
vector components in (\ref{7}) we obtain
 \beax
 \lefteqn{\frac{1}{2} \r_n(B_1+iB_2) w_{n+1} (z q^{-y})^{j-n-1} + n
 B_3 w_n (zq^{-y})^{j-n}}
 \\ &&+\frac{1}{2} \r_{n-1}(B_1-iB_2) w_{n-1} (z
 q^{-y})^{j-n+1}\hspace{2cm}
 \\
 &=&j w_n (zq^{-y})^{j-n},
 \eeax
with $\r_n= \sqrt{j(j+1)-n-n^2}$, and for $|n|\le j$. This leads
to the following equation:
 \beax\lefteqn{\frac{1}{2} \r_n  w_{n+1}
 (1-B_3^2) + \frac{1}{2} \r_{n-1} w_{n-1} (1+B_3)^2 }
 \\
 &&-nB_3 w_n (1+B_3)
 \\
 &=&j w_n (1+B_3).\hspace{3.5cm}
 \eeax
By a straightforward calculations one verifies that
 \beax \frac{1}{2} \r_n w_{n+1} + \frac{1}{2} \r_{n-1} w_{n-1}  &=& j w_n,\\
      \r_{n-1}  w_{n-1} &= &(j+n) w_n,\\
      -\frac{1}{2} \r_n w_{n+1} + \frac{1}{2} \r_{n-1} w_{n-1} &=&n w_n.
 \eeax
This proves (\ref{7}).
\end{proof}

\section{Explicit diagonalizations of small-site spin $1/2$ Hamiltonians}

\subsection{$H^{+-}_{12}(\vec{B})$}

Here we diagonalize the two-site Hamiltonian,
$H^{+-}_{12}(\vec{B})$, with magnetic field not parallel to
$z$-axis at $y=1$. By the {\sf XX} symmetry we may assume $B_2=0$.
Since we already know two eigenvalues, namely,
$\pm\frac{1}{2}\sqrt{B_1^2+B_3^2}$, it is best to factor them out
from the characteristic equation. Another way is to diagonalize
the Hamiltonian restricted to the orthogonal complement of the two
known eigenvectors. The Hamiltonian is of the form
 \beax \lefteqn{2 H^{+-}_{[1,2]}(B_1,B_3)}
 \\
 &=&\left(\begin{array}{cccc} B_3&0          &B_1        &0\\
                             0  &1-A+B_3&-\D^{-1}   &B_1\\
                 B_1&-\D^{-1}   &1+A-B_3&0\\
                 0  &B_1        &0          &-B_3
     \end{array}
   \right) .
 \eeax
The characteristic polynomial, $p$, is equal to
 \beax p(t)
 &=&(t^2-B_3^2)(t^2-2t+2AB_3-2B_1^2-B_3^2)
 \\
 &+&2B_1^2(t-AB_3) + B_1^4.
 \eeax
We divide this polynomial by $t^2-B_1^2-B_3^2$ (Note that we have
multiplied the Hamiltonian by two) obtaining
$$p(t)= (t^2 -2t -B_1^2 - B_3^2 +2B_3 A)(t^2-B_1^2-B_3^2).$$
The two eigenvalues we are looking for are thus
$$t_\pm = 1\pm \sqrt{1 + B_1^2 + B_3^2 - 2 B_3 A}.
$$
One can easily verify that
 $$
 \sqrt{B_1^2+B_3^2}\ge1-\sqrt{1+B_1^2+B_3^2-2B_3A}.
 $$
Hence, the gap between the lowest eigenvalues of
$H^{+-}_{[1,2]}(B_1,B_2,B_3)$ is
equal to
 \be\label{kinkgap}
   g_2^{+-}(\vec{B}) = \mfr{1}{2}  - \mfr{1}{2} \sqrt{1 + \|\vec{B}\|^2 - 2B_3A}
                  +\mfr{1}{2} \|\vec{B}\|,
 \ee
which is a positive function.

\subsection{$H^{++}_{[12]}(\vec{B}), B_1^2+B_2^2>0$}

The diagonalization of the two-site droplet Hamiltonian,
$H^{++}_{[12]}(\vec{B})$,
with the field at $y=1$ is very similar to the two-site kink Hamiltonian.
We have
 \beax
 \lefteqn{2 H^{++}_{[12]}(B_1,B_3) -A{\bf 1}}
 \\
 &=&\left(\begin{array}{cccc} B_3-A&0       &B_1        &0\\
                             0        &1+B_3   &-\D^{-1}   &B_1\\
                 B_1      &-\D^{-1}&1-B_3      &0\\
                 0        &B_1     &0          &A-B_3
     \end{array}
   \right) .
 \eeax
The characteristic polynomial, $q$, of the rhs is equal to
 \beax
 q(t)
 &=&(t^2-(B_3-A)^2)(t^2 - 2t - 2B_1^2 -B_3^2 + A^2)
 \\
 &-&2B_1^2(t-A^2) + B_1^4.
 \eeax
$t=\pm \sqrt{B_1^2+(B_3-A)^2}$ are two roots and we factor them out
from $q(t)$, and obtain
 \beax
 q(t)&=&(t^2 - 2t - B_1^2 - B_3^2 + A^2) (t^2-B_1^2-(B_3-A^2)).
 \eeax
The two new eigenvalues of $H^{++}_{[12]}(\vec{B})$ are
 $$
 t_\pm =  \mfr{1}{2} \left(1 \pm \sqrt{1 + \|\vec{B}\|^2 -A^2} +A\right) .
 $$
The gap above the ground state is therefore
 \bea \label{dropletgap1}
 g^{++}_2(\vec{B},\D) &= &
     \mfr{1}{2}\left(1 - \sqrt{1 + \|\vec{B}\|^2 -A^2} \right.
 \nonumber
 \\
 &&+\left.\sqrt{B_1^2 +B_2^2 +(B_3-A)^2}\right).
 \eea

\subsection{$H^{++}_{[13]}(0,0,B), B<A$}

Since in this case there is only a magnetic field in the
$z$-direction we can easily diagonalize the three-site droplet
Hamiltonian, $H=H^{++}_{[13]}(0,0,B)$, with the field in the
middle at $y=2$. Then $H$ commutes with the symmetry $S:S(u_{\s_1}
\otimes u_{\s_2} \otimes u_{\s_3}) = u_{\s_3}\otimes
u_{\s_2}\otimes u_{\s_1}$, where $\sigma_i=\pm$. We choose the
following eigenbasis of $S$:
 \beax v_1&=&(1,0)\otimes(1,0)\otimes(1,0),
 \\
 v_2&=&(0,1)\otimes(0,1)\otimes(0,1),
 \\
 v_3&=&\mfr{1}{\sqrt{2}}(1,0)\otimes(1,0)\otimes(0,1) - \mfr{1}{\sqrt{2}}
       (0,1)\otimes(1,0)\otimes(1,0),
 \\
 v_4&=&\mfr{1}{\sqrt{2}}(1,0)\otimes(0,1)\otimes(0,1) -
       \mfr{1}{\sqrt{2}}(0,1)\otimes(0,1)\otimes(1,0),
 \\
 v_5&=&\mfr{1}{\sqrt{2}}(1,0)\otimes(1,0)\otimes(0,1) +\mfr{1}{\sqrt{2}}
       (0,1)\otimes(1,0)\otimes(1,0),
 \\
 v_6&=&(1,0)\otimes(0,1)\otimes(1,0),
 \\
 v_7&=&(0,1)\otimes(1,0)\otimes(0,1) ,
 \\
 v_8&=&\mfr{1}{\sqrt{2}}(0,1)\otimes(0,1)\otimes(1,0)  + \mfr{1}{\sqrt{2}}
       (1,0)\otimes(0,1)\otimes(0,1).
 \eeax
$v_1,v_2,v_3,v_4$ are eigenvectors of $H$ with eigenvalues
$e_1=\frac{B}{2}$, $e_2=A-\frac{B}{2},e_3=\frac{1}{2}(A+1+B)$ and
$e_4=\frac{1}{2}(A+1-B)$, respectively. What remains
are two copies of the two-dimensional matrix (due to the symmetry $S$)
 \beax N(B)&=&\mfr{1}{2}\left(\begin{array}{cc}A+1+B&-(\sqrt{2}\D)^{-1}
             \\-(\sqrt{2}\D)^{-1}&A+1-B\end{array}\right).
 \eeax
The matrix $N$ is equal to $H^{++}(B)$ reduced to the span$\{v_5,v_6\}$,
as well as
to span$\{v_7,v_8\}$. The eigenvalues are equal to
 \bea
 e_5 = e_7&=& \mfr{1}{2}\left(A+1 - \sqrt{\mfr{1}{2}\D^{-2} + B^2}\right),
 \\
 e_6 = e_8&=& \mfr{1}{2}\left(A+1 + \sqrt{\mfr{1}{2}\D^{-2} + B^2}\right).
 \eea
Notice that for $B<\bar{B}=\frac{3A^2-4A+1}{4(1-A)}$,
 \be
 e_1 < e_5 = e_7 < e_2.
 \ee
This says that the lowest energies in the total $S^3$-sectors are
ordered (though not strictly) by their energy. The gap for $B\le
A$ is equal to
 \bea \label{dropletgap2}
 \lefteqn{g^{+}_3(B,\D)}
 \\
 && = \left\{\begin{array}{lcl}\mfr{1}{2}
 \left(1+A-\sqrt{\mfr{1}{2}\D^{-2} + B^2} -
 B \right)&\mbox{ for }&B\le \bar{B}\\
             A-B&\mbox{ for }& \bar{B}\le B\le A
             \end{array}\right..
 \nonumber
 \eea

\section{Excitation ${\cal E}_-(B)$}

Here we calculate the lowest eigenvalue of $\tilde{H}(B)=H_0^{++}
+ B (S_0^3-\frac{1}{2})$ in the sector with one overturned spin.
Since we want to avoid complications from finite chain boundary
effects, we prefer to treat the infinite volume case with the
magnetic field at 0, say. We first take $B\ge0$.

Let $\psi=\sum_x a_x S_x^-|\!\!\uparrow\rangle$. Then, for $\psi$
being an eigenvector of $\tilde{H}(B)$ with energy $\tilde{\cal
E}$, we have $\langle\uparrow\!| S_x^+ \tilde{H}(B)|\psi\rangle
=\tilde{{\cal E}} a_x$, and thus we get the equations
 \bea
 a_{x+1}&=&2\D (1-{\cal E}) a_x -a_{x-1}, \quad\mbox{for}\quad
 |x|>1
 \label{first}\\
 a_1&=&2\D(1-\tilde{{\cal E}} + B)a_0 -a_{-1} .
 \label{second}
 \eea
It turns out that in addition to the pure absolutely continuous
spectrum of the discrete Laplacian (in the units here, it is the
interval $[1-\D^{-1},1+\D^{-1}]$) there are two (a highest and a
lowest) eigenvalue generated by the perturbation $B S_0^3$. Let
 $$
 r_\pm=\D(1-\tilde{{\cal E}}) \pm\sqrt{\D^2(1-\tilde{{\cal E}})^2-1}
 $$
be the solutions to the characteristic polynomial. Then, all
solutions of (\ref{first}) are of the form $a_x = \a_1 r^x + \a_2
r^{-x}$ for $|x|>1$. Notice that $r_+=r_-^{-1}$. We now look for
the solution $a_x=r^{|x|}$ with $r=r_-$ which produces an
eigenvector. With this choice, we have $|r_-|<1$, we insert this
into (\ref{second}). Then we get
 $$
 \D B=\sqrt{\D^2(1-\tilde{{\cal E}})^2-1},
 $$
from which we conclude $\tilde{{\cal E}}_\pm(B) =
1\pm\sqrt{B^2+\D^{-2}}$. From the gap at $B=0$ we know \cite{KN1}
that $\tilde{{\cal E}}(0) = 1-\D^{-1}$. Thus, the correct solution
is $\tilde{{\cal E}}_-(B)$ which, for the original Hamiltonian of
interest, namely $H_0^{++} + B S_0^3$, has to be shifted back by
$B/2$.

Similarly, if $B\le 0$, then we study $\tilde{H}(B) = H_0^{++} + B
(S_0^3 + \frac{1}{2})$ which amounts to replacing $B$ by $-B$ in
(\ref{first}-\ref{second}).

The lowest energy state of $H_0^{++} + B S_0^3$ in the sector with one
overturned spin is thus
 \be \label{excit}
 {\cal E}_-(B) = 1-\sqrt{B^2+\D^{-2}} +\frac{1}{2}|B|.
 \ee
\end{appendix}

\begin{acknowledgments}
This material is based upon work supported by the National Science
Foundation under Grant \# DMS-0070774. B.N. would like to thank
the Dipartimento de Matematica of the Universit\`a de Bologna,
where this work was initiated, for warm hospitality. We are
indepted to Daniel Ueltschi for a very thorough reading of the
manuscript and the figure in the introduction. W.S. wants to thank
Shannon Starr for providing the matlab programs \cite{St2} with
which many of the propositions in this article where tested, and
dedicates this paper to the memory of his uncle Ernst.
\end{acknowledgments}


%
\end{document}